\documentclass[pdf,twocolumn,superscriptaddress,floatfix,showpacs,amsmath,longbibliography,nofootinbib,amssymb]{revtex4-1}

\usepackage[colorlinks=true,citecolor=blue,linkcolor=blue]{hyperref}
\usepackage[usenames]{color}
\usepackage{subfigure}
\usepackage{dcolumn}
\usepackage{mathrsfs}
\usepackage{bm}
\usepackage{amsmath,amssymb,epsfig,float,graphics}

\usepackage{braket}
\usepackage{physics}

\begin{document}

\title{Quantum Simulation of Oscillatory Unruh Effect with Superposed Trajectories}

\author{Xu Cheng}\email{Equal contribution.}
\affiliation{Laboratory of Spin Magnetic Resonance, School of Physical Sciences, Anhui Province Key Laboratory of Scientific Instrument Development and Application, University of Science and Technology of China, Hefei 230026, China}
\affiliation{Hefei National Laboratory, University of Science and Technology of China, Hefei 230088, China}

\author{Yue Li}\email{Equal contribution.}
\affiliation{Laboratory of Spin Magnetic Resonance, School of Physical Sciences, Anhui Province Key Laboratory of Scientific Instrument Development and Application, University of Science and Technology of China, Hefei 230026, China}

\author{Zehua Tian}\email{tianzh@ustc.edu.cn}
\affiliation{School of Physics, Hangzhou Normal University, Hangzhou, Zhejiang 311121, China}
\affiliation{Laboratory of Spin Magnetic Resonance, School of Physical Sciences, Anhui Province Key Laboratory of Scientific Instrument Development and Application, University of Science and Technology of China, Hefei 230026, China}

\author{Xingyu Zhao}
\affiliation{Laboratory of Spin Magnetic Resonance, School of Physical Sciences, Anhui Province Key Laboratory of Scientific Instrument Development and Application, University of Science and Technology of China, Hefei 230026, China}

\author{Xi Qin}
\affiliation{Laboratory of Spin Magnetic Resonance, School of Physical Sciences, Anhui Province Key Laboratory of Scientific Instrument Development and Application, University of Science and Technology of China, Hefei 230026, China}
\affiliation{Hefei National Laboratory, University of Science and Technology of China, Hefei 230088, China}

\author{Yiheng Lin} \email{yiheng@ustc.edu.cn}
\affiliation{Laboratory of Spin Magnetic Resonance, School of Physical Sciences, Anhui Province Key Laboratory of Scientific Instrument Development and Application, University of Science and Technology of China, Hefei 230026, China}
\affiliation{Hefei National Research Center for Physical Sciences at the Microscale, Hefei 230026, China}
\affiliation{Hefei National Laboratory, University of Science and Technology of China, Hefei 230088, China}

\begin{abstract}

The Unruh effect predicts an astonishing phenomenon that an accelerated detector would detect counts despite being in a quantum field vacuum in the rest frame. 
Since the required detector acceleration for its direct observation is prohibitively large,  recent analog studies on quantum simulation platforms help to reveal various properties of the Unruh effect and explore the not-yet-understood physics of quantum gravity. 
To further reveal the quantum aspect of the Unruh effect, analogous experimental exploration of the correlation between the detector and the field, and the consequences for coherent quantum trajectories of the detector without 
classical counterparts, are essential steps but are currently missing. Here, we utilize a laser-controlled trapped ion to experimentally simulate an oscillating detector coupled with a cavity field. We observe joint excitation of both the detector and the field in the detector's frame, coincide with the coordinated dynamics predicted by the Unruh effect. Particularly, we simulate the detector moving in single and superposed quantum trajectories, where the latter case shows coherent interference of excitation. Our demonstration reveals properties of quantum coherent superposition of accelerating trajectories associated with quantum gravity theories that have no classical counterparts, and 
may offer a new avenue to investigate phenomena in quantum field theory and quantum gravity. We also show how a generalization of the method and results in this work may be beneficial for direct observation of the Unruh effect.
\\           \par \
\end{abstract}
\maketitle
\baselineskip=0.45 cm

\newpage

\section{Introduction}\label{sec1}
The exploration of quantum gravity is one of the most exciting areas in modern physics, with quantum mechanics applied to gravitational systems or curved spacetimes. 
Relativistic quantum field theory (QFT) predicts that various quantum constructs such as particle content and vacuum fluctuations of the quantum field are observer-dependent. A prominent paradigm is the Unruh effect \cite{PhysRevD.14.870, RevModPhys.80.787}, where a pointlike two-level Unruh-DeWitt detector~\cite{PhysRevD.14.870, RevModPhys.80.787, PhysRevD.29.1047} accelerating in Minkowski vacuum would experience a non-vacuum field, leading to detection events. 
Despite a number of theoretical advancements, involving for example Berry phase \cite{PhysRevLett.107.131301, PhysRevLett.129.160401} and cavity resonance \cite{PhysRevLett.91.243004, PhysRevLett.125.241301, PhysRevLett.129.111303}, direct experimental verification of the Unruh effect remains challenging given large accelerations for any appreciable non-inertial effects.
Bridging the gap between the theoretical predictions and experimental verification, artificial analogue gravity systems provide a new avenue to study relativistic QFT effects~\cite{AG, RevModPhys.84.1, RevModPhys.86.153}. As such, proof-of-principle experimental demonstration from the perspective of quantum simulation could be within reach. 
Representative paradigms include analogue Hawking radiation \cite{doi:10.1126/science.1153625, PhysRevLett.106.021302, PhysRevLett.105.203901, PhysRevLett.117.121301, PhysRevLett.124.141101, PhysRevLett.123.161302, PhysRevLett.122.010404, AHR2, AHR, NatCommun.14.3263}, cosmological particle production \cite{doi:10.1126/science.1237557, PhysRevX.8.021021, PhysRevLett.123.180502, PhysRevLett.128.090401, CPC, CPC2}, and the dynamical Casimir effect \cite{PhysRevLett.105.233907, PhysRevLett.109.220401, doi:10.1073/pnas.1212705110, DCE, PhysRevLett.124.140503}. Recently, degradation of quantum correlation due to the Unruh effect has been experimentally simulated \cite{Unruh-effect,PhysRevResearch.6.013233}. In addition, Unruh radiation has been reported through Bose-Einstein condensate \cite{SUE}, where a dynamical two-mode squeezed quantum field is constructed and observed, in analogy to the field seen by the non-inertial observer mathematically. 

A significant step closer to the Unruh's original proposal is to directly simulate the behavior of the non-inertial detector~\cite{PhysRevLett.101.110402, PhysRevResearch.2.042009, PhysRevLett.125.213603, PhysRevD.106.L061701, Tian-EPJC, PhysRevLett.126.117401} and its interplay with the quantum field~\cite{PhysRevB.92.064501,NewJ.Phys.22.033026}. This provides a parallel and mutually authenticated avenue to the Unruh effect demonstration. 
However, these aspects remain to be pursuit experimentally. Such a demonstration may also provide valuable insight into its real-life observation.

Beyond the original Unruh effect, the oscillatory Unruh effect has been attracting increasing attention~\cite{Oscillatory-Unruh, PhysRevA.99.053833}. The trajectory in the original Unruh effect proposal is a uniformly accelerated motion, which makes the detector perceive a thermal bath rather than a vacuum state. However, this demands an extremely high acceleration to be maintained over extensive distances, posing significant practical challenges. To avoid these difficulties, the uniformly accelerated motion could be substituted with bounded trajectories that involve acceleration. Typical examples include uniform circular motion~\cite{PhysRevResearch.2.042009} and simple harmonic oscillation~\cite{Oscillatory-Unruh, PhysRevA.99.053833}. Even with the altered trajectory forms, vacuum excitation phenomena still emerge in the presence of acceleration, but not in the original manifestation as a thermal bath. Furthermore, cavities are introduced to enhance the coupling between the Unruh-DeWitt detector and the vacuum. The simulation scheme involving an oscillatory trajectory in a cavity has been proposed in the superconducting system~\cite{PhysRevB.92.064501}. The quantum simulation of the oscillatory Unruh effect is beneficial for examining the evolution and the form of excitation, paving ways to the direct real-life observation in the future. The proposal of direct real-life observation is discussed in Sec. \ref{sec6}.

In addition, the Unruh-DeWitt detector model could be extended to travel in \emph{quantum superpositions of classical trajectories} \cite{PhysRevD.102.045002, PhysRevD.102.085013, PhysRevLett.125.131602} and \emph{of spacetime} \cite{PhysRevLett.129.181301, PhysRevD.107.045014}, resulting in possible exotic interference for the detector dynamics. Such ``bottom-up" approach may provide an essential step toward studying quantum-gravitational physics, for example the fundamental physics in quantum reference frames \cite{QRF, PhysRevD.105.125001, Giacomini2021spacetimequantum}, quantum causal structures \cite{QCS1, QCS2, 10.3389/fphy.2020.525333,  PhysRevLett.125.131602}, and novel quantum gravity effects upon quantum matter \cite{PhysRevLett.119.240402, PhysRevLett.119.240401, SCPMA.67.}, which have no classical counterparts. Conversely, this may also help further explore a new possible witness for quantum gravity rather than directly detecting gravitons, or detecting quantum gravitational vacuum fluctuations in the future. 
However, experimental quantum simulations for the above-mentioned effect remain an open question. Such demonstration would help understanding those operational methods for constructing and quantitatively analyzing effects produced by general spacetime and motion-trajectories superpositions.

In this work, we use a vibrating trapped ion to experimentally simulate the Unruh effect, observing the coordinated dynamics of both the detector and quantum field, and exploring the exotic phenomena induced by superposed trajectories. 
Firstly, the detector motion is modeled with a single oscillatory trajectory \cite{Oscillatory-Unruh, PhysRevA.99.053833}, spatially confined and thus favorable for experimental implementation. We experimentally observe the corresponding analogue photon creation out of the vacuum along with a simultaneous detector excitation. Secondly, we consider superposed trajectories for the detector motion, as controlled by additional spin levels of the same single ion. Such implementation provides a simplification compared to the previous proposal utilizing an additional physical spin and control operations across spins \cite{PhysRevD.102.085013}. We show the coherent interference between trajectories, and the corresponding novel non-classical excitation dynamics for the non-inertial detector. Our result supports the Unruh's prediction regarding particle creation by non-inertial motion, as well as the novel coherent effects in quantum field theory relating to quantum gravity. Our proof-of-principle simulation may provide a possible path towards direct observation of the Unruh effect with all the quantum elements involved, as discussed at the end of this work for potential experimental implementation.

\section{Models}
\subsection{Theoretical model}\label{sec2}

\begin{figure}
\centering
\includegraphics[width=\columnwidth]{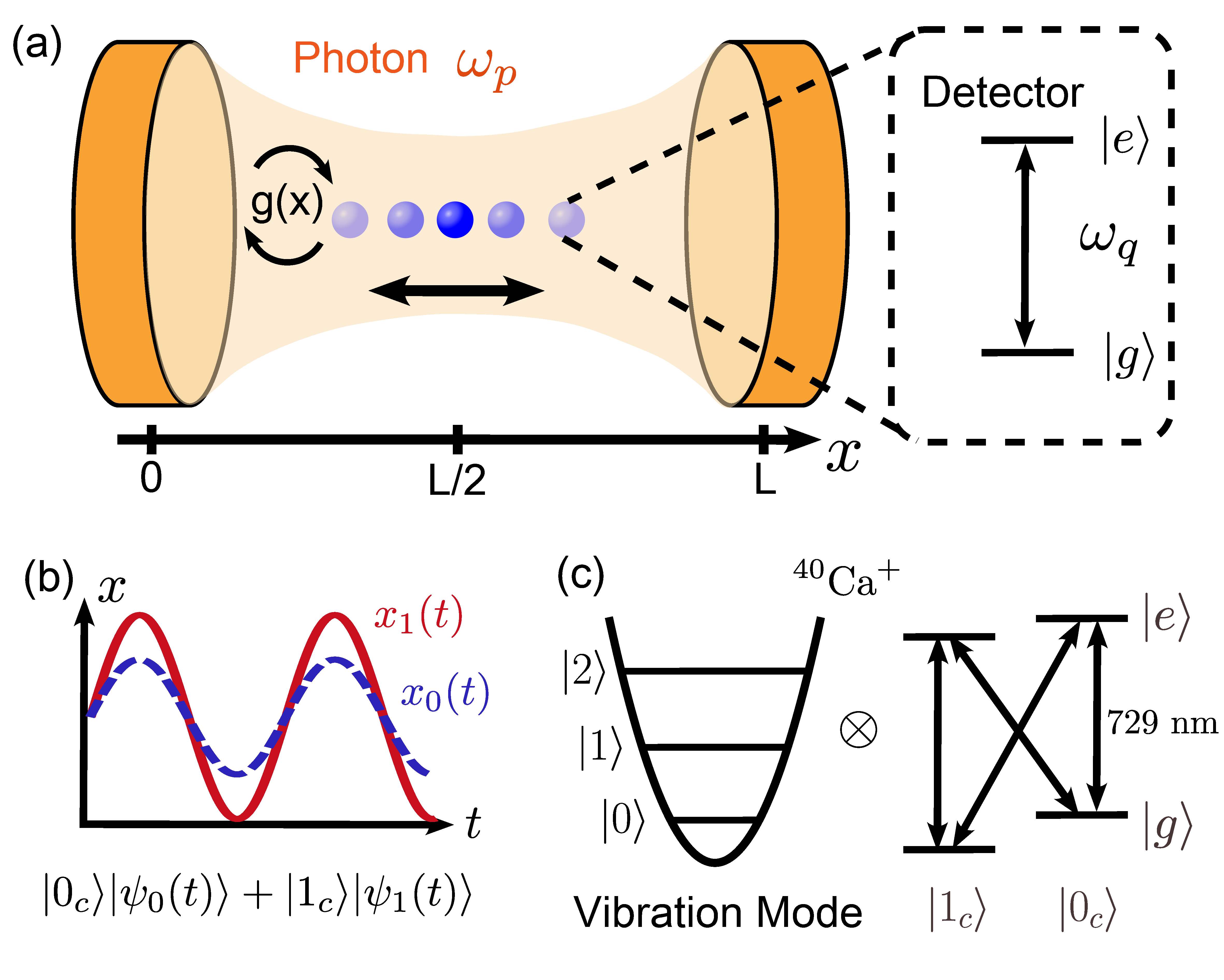}
\caption{
Illustration of the model and the experimental implementation. 
(a) The detector oscillating in a cavity of length $L$. The energy levels of the detector and photon of cavity mode are $\omega_q$ and $\omega_p$. The coupling strength $g(x)$ between the detector and the photon is position-dependent due to the stationary wave form in the cavity.
(b) The quantum superposition of two trajectories $x_0(t)$ and $x_1(t)$. Trajectory $x_0(t)$ ($x_1(t)$) is labeled with the solid red line (dashed blue line).  An additional control qubit $\ket{i_c}$ is introduced, with $i=\{0,1\}$, designating individual paths for the detector.  
(c) Experimental scheme for quantum simulation of the detector-photon model with superposed trajectories. Four combined states of detector and control qubit are mapped to four internal levels of a $^{40}\rm{Ca}^+$ ion. Transitions between $\ket{g}$ and $\ket{e}$ could be individually driven with $729\text{ nm}$ laser beams in different frequencies. 
For single trajectory simulation, only the energy levels and transition with respect to $\ket{0_c}$ are utilized.}
\label{fig:model}
\end{figure}

As illustrated in Fig. \ref{fig:model} (a), our model consists of a moving detector and photon modes in a cavity. Without loss of generality, the detector is modeled as being coupled to the resonant photon mode only. Thus, the detector-photon system can be described by a Rabi model \cite{Phys.Rev..49.324,Phys.Rev..51.652}. Specifically, we consider a two-level detector described by the Pauli operators $\sigma_x,\sigma_y$ and $\sigma_z$, with an energy spacing $\hbar\omega_q$. The ground and excited states of the detector are respectively denoted as $\ket{g}$ and $\ket{e}$. 
\emph{In the detector's frame}, the standing-wave photon mode in the cavity is considered with the energy spacing $\hbar\omega_p=\hbar kc$ and the wave vector $k=2\times 2\pi/L$, where $L$ is the cavity length and $c$ is the speed of light. 
The photon number operator is $N=a^\dagger a$ with the eigenstates denoted as $\ket{n}$, where $a$ and $a^\dagger$ are the corresponding annihilation and creation operators. 
Following previous proposals for oscillatory trajectories\cite{ NewJ.Phys.22.033026,PhysRevB.92.064501}, the coupling strength $g(x(t))=g_0\sin(kx(t))$ would have a periodic space modulation due to the sinusoidal field distribution of the standing wave, which depends on the detector's trajectory $x(t)$. 
The Hamiltonian of the system reads
\begin{equation}\label{Eq:H_single}
    H = \hbar\omega_p a^\dagger a + \cfrac{\hbar\omega_q}{2}\sigma_z +\hbar g_0 \sin(k x(t))\sigma_x(a+a^\dagger),
\end{equation}
where the last term denotes the detector-photon interaction. The term $x(t)$ describes the detector trajectory dependent on time. Thus, the excitation dynamics for both the detector and the photon field is determined by $x(t)$, via the effective coupling strength $\hbar g_0\sin(k x(t))$. 
The coupling strength is mathematically equivalent to a temporal modulation form $g^\prime(t):=g_0\sin(k x(t))=g(x(t))$. We utilize this correspondence to perform the quantum simulation and examine the evolution under Eq.\ (\ref{Eq:H_single}). 
Consider a non-inertial oscillatory trajectory $x(t)=\bar{x}+A\sin(\omega t)$, where $\bar{x}\in[0,L]$ is the initial position, $A$ is the amplitude and $\omega$ is the oscillation frequency of the trajectory. With proper choice of parameters, a non-zero acceleration $a(t)={\rm d}^2 x(t)/{\rm d}t^2$ excites the detector and photon field from the initial ground states $\ket{g,n=0}$, leading to a joint detector-photon state $\ket{\psi(t)}$. After a dynamics of interaction duration $t$, detector excitation $\langle\sigma_z\rangle>-1$ and photon creation $\langle N\rangle>0$ could be observed. However, we expect that negligible excitation of the detector or photon field could occur for a static trajectory $x(t)=\bar{x}$ or an inertial trajectory $x(t)=v t$ with the detector velocity $v\leq~c$. Interestingly, for the latter case with inertial motion, theoretical analysis \cite{PhysRevLett.128.163603} predicts that resonant excitation for both the detector and photon only occurs when $\omega_q+\gamma(ck-\vec{k}\cdot\vec{v})=0$, where $\gamma=1/\sqrt{1-v^2/c^2}$ is the Lorentz factor, corresponding to a speed of $v=c(1+\omega_q/\gamma\omega_p)$, which is superluminal. In comparison, oscillatory trajectories would induce detectable excitation without a relativistic treatment~\cite{NewJ.Phys.22.033026, PhysRevB.92.064501}. 


Inspired by the quantum superposition of spacetime \cite{PhysRevD.102.045002, PhysRevD.102.085013, PhysRevLett.125.131602, PhysRevLett.129.181301, PhysRevD.107.045014}, we extend the above single trajectory model to multiple coherent trajectories. To initialize the detector in a trajectory superposition illustrated in Fig. \ref{fig:model} (b), we introduce an additional control degree of freedom, $|i_c\rangle$ designating the individual trajectories $x_{i}(t)$ for the detector, with $i=\{0, 1\}$. 
Combining two-level detector and the control degree of freedom, a four-level system, labeled as $\{\ket{0_c},\ket{1_c}\}\otimes\{\ket{g},\ket{e}\}$, could be used to describe the system. This four-level map enables the simulation of superposed trajectories on a multi-level trapped ion. 
Following Eq. (\ref{Eq:H_single}), we introduce Hamiltonian for the model of superposed trajectories as:
\begin{equation}\label{Eq:sup_total}
\begin{aligned}
    H =&\hbar\omega_p a^\dagger a + \cfrac{\hbar\omega_q}{2}\sigma_z\\
     + \sum_{i=0,1} &\hbar g_0 \sin(k x_i(t))\ketbra{i_c}\otimes\sigma_x(a+a^\dagger).
\end{aligned}
\end{equation}
For each of the trajectories, the dynamics after a duration $t$ gives a joint detector-photon state $\ket{\psi_{i=0,1}(t)}$,  similar to the case of single trajectory. Thus, an initial state of the system $\frac{1}{\sqrt{2}}(\ket{0_c}+\ket{1_c})\ket{g, n=0}$ would evolve to $\ket{\Psi(t)}=\frac{1}{\sqrt{2}}(\ket{0_c}\ket{\psi_0(t)}+\ket{1_c}\ket{\psi_1(t)})$ after a duration $t$, where a superposed control qubit leads to superposed trajectories for the dynamics of the detector and photon field. Following the single trajectory model, the relativistic effect could be validated if detector excitation $\langle\ketbra{i_c}\otimes\sigma_z\rangle > \langle\ketbra{i_c}\rangle\times(-1)$ and photon creation $\langle \ketbra{i_c}\otimes N\rangle>0$, with $i=0,1$. Explicitly,
the final state $\ket{\Psi(t)}$ can be rewritten as 

\begin{eqnarray}\label{eq:PsiT}
\nonumber\ket{\Psi(t)}&=&\frac{1}{2}\bigg[\ket{+_c}(\ket{\psi_0(t)}+\ket{\psi_1(t)})\\
&&+\ket{-_c}(\ket{\psi_0(t)}-\ket{\psi_1(t)})\bigg].
\end{eqnarray}

Thus, measurements of $\langle\ketbra{\pm_c}\otimes\sigma_z\rangle$ and $\langle\ketbra{\pm_c}\otimes N\rangle$, with  $\ket{\pm_c}=(\ket{0_c}\pm\ket{1_c})/\sqrt{2}$, explicitly give the expectation of $\langle\sigma_z\rangle$ and $\langle N\rangle$ with respect to the superposed states $\ket{\psi_0(t)}\pm\ket{\psi_1(t)}$ up to a normalization factor. These results correspond to the quantum coherence between the trajectories, revealing a combination of relativistic and quantum-mechanical effects. The extended model could reveal novel predictions in relation to the physics of quantum gravity.

\subsection{Experimental setup}\label{sec3}

To experimentally simulate the oscillatory Unruh effect, we utilize the unique features for trapped ions in preparation, control, and detection of internal spin and external vibration states \cite{RevModPhys.75.281, RevModPhys.85.1103, PhysRevLett.100.200502, doi:10.1126/science.1261033, TIE}, which has brought successful simulation of a variety of fundamental physics such as relativistic Dirac equation \cite{Dirac, PhysRevLett.106.060503, PhysRevLett.128.200502}, various spins models \cite{SModle, Schneider_2012, RevModPhys.93.025001}, and quantum field theories \cite{QFT}. 
We develop techniques to control the coupling between the multiple levels and external vibration, which enables coherent interference between trajectories. 
We trap a $^{40}\text{Ca}^+$ ion in a linear Paul trap and an ambient magnetic field of $0.69$ mT. As depicted in Fig. \ref{fig:model} (c), we utilize four fine structure sub-levels of a $^{40}\text{Ca}^+$ ion to simulate the internal levels of the detector. For simulation of superposed trajectories, we map $\ket{0_c,g}$, $\ket{1_c,g}$, $\ket{0_c,e}$, $\ket{1_c,e}$ states to fine structure sub-levels $\ket{S_{1/2},+1/2}$, $\ket{S_{1/2},-1/2}$, $\ket{D_{5/2},+1/2}$, $\ket{D_{5/2},-1/2}$, respectively. 
For the case of a single trajectory, only the pair of energy levels with respect to $i=0$ and the corresponding transition are utilized, as shown in Fig.~\ref{fig:model}~(c). 
The photon field is simulated by the quantized ion vibration (i.e. phonons) with mode frequency $\omega_r$. 
Coupling between the detector and the photon could be achieved with multi-tune coherent 729 nm laser pulses. The laser detuning is set to be $-\omega_q\pm(\omega_r-\omega_p)$. The Hamiltonian Eq. (\ref{Eq:H_single}) could be derived in the interaction picture with proper rotation-wave-approximation.
Coupling modulation induced by the trajectory $g(x(t))$ reflects in temporal modulation of strength of the laser pulses $g^{\prime}(t)=g(x(t))$ in the experiment. The experimental modulation is fully programmable for general trajectory forms. After evolution, the combined final state is measured with additional process and detection. Details of the experiment scheme and the analysis of the theory above are explained in (See in Supplement Material).

\section{Results}
\subsection{Experimental simulation of single trajectory}\label{sec4}

First, we simulate the oscillatory Unruh effect of a single trajectory. 
The trajectory is set as a sinusoidal form of $x(t)=\bar{x}+A\sin(\omega t)$. {Here the trajectory is set to start from the center of the cavity, i.e. $\bar{x}=L/2$. }Then the modulated coupling strength becomes $g_0\sin(kx(t))=g_0\sin(\bar{u}+u\sin(\omega t))$, where dimensionless quantities are introduced to describe the detector position {$\bar{u}=4\pi \bar{x}/L=2\pi$} and amplitude $u = 4\pi A/L$. 
Here we demonstrate the case with $\omega = \omega_p + \omega_q$ \cite{PhysRevB.92.064501} so that the modulated coupling resonantly excites both the detector and the photon field. 
This embodies that the coupling is equivalent to the form $i\sigma_+a^\dagger -i \sigma_- a$ under Floquet analysis \cite{PhysRevLett.128.163603} (See in Supplement Material) given $g_0\ll\{\omega_p,\omega_q\}$ considered in this work. This is remarkable since such coupling plays a significant role only with the presence of non-inertial motion, corresponding to the physics of relativity. In this implementation, the periodic motion generates a series of resonances, enabling the resonant excitation of both the photon field and detector. 
Therefore, the evolution could, as a result of the detector's oscillatory motion, arise coupling between $\ket{g, n=0}$ and $\ket{e, n=1}$. So the measured values of $\langle\sigma_z\rangle$ and $\langle N \rangle$ oscillate in a sinusoidal pattern. 
For experimental implementation, we set $\omega_p = \omega_q=2\pi\times150~\rm{kHz}$ and $g_0=2\pi\times 1.92$~kHz. The initial state is the ground state $\ket{g, n=0}$. 
Two different oscillation amplitudes of the trajectory with $u = 0$ and $1.5$ are demonstrated.


\begin{figure}
\centering
\includegraphics[width=1.0\columnwidth]{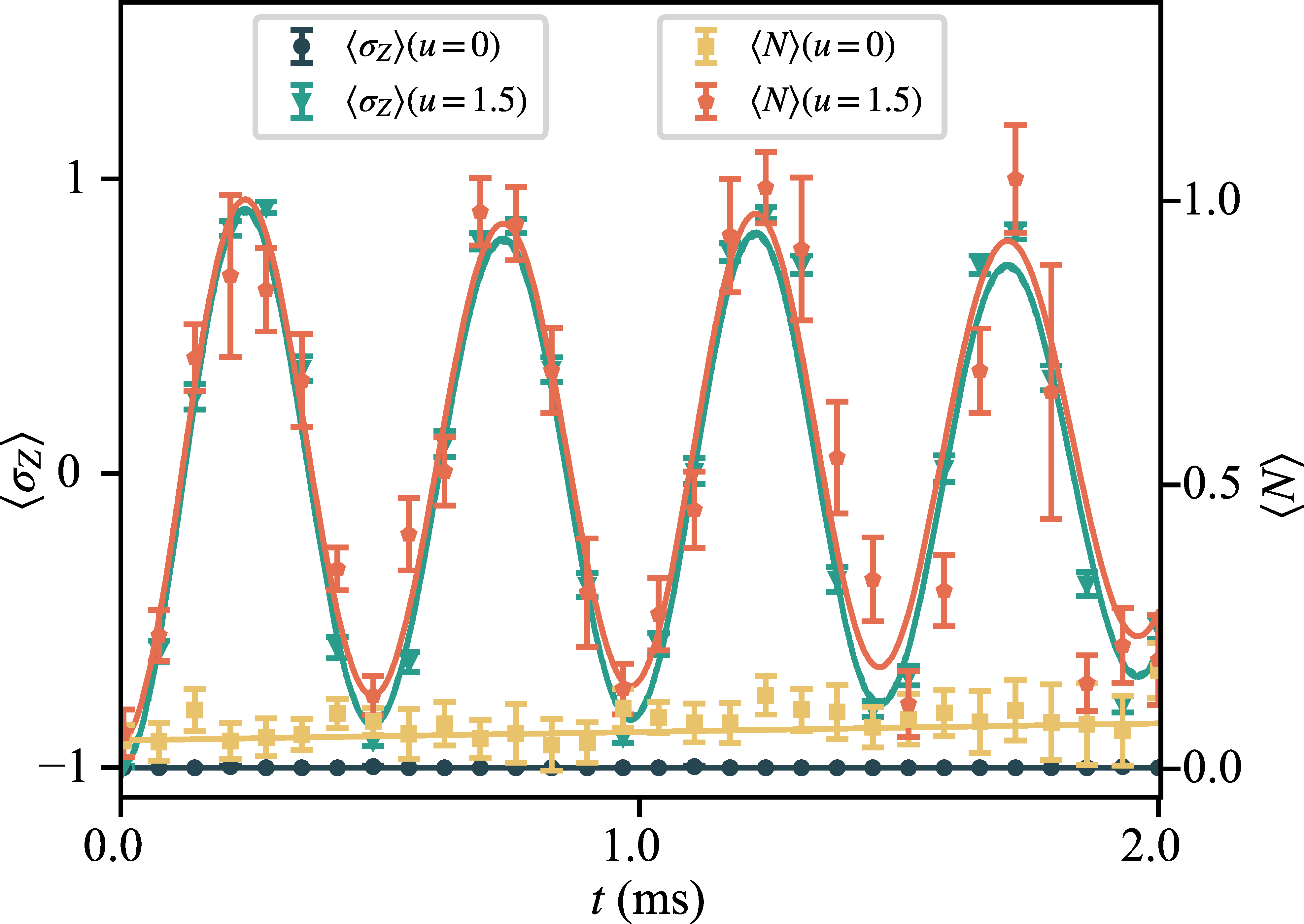}
\caption{Experimental simulation of oscillatory Unruh effect with single trajectory. The evolution of detector excitation $\langle\sigma_z\rangle$ and photon creation $\langle N \rangle$ for different oscillation amplitudes. For $u=0$ ($u=1.5$), data points with error bars represent the experimental results, with dark green dots (light green triangles), yellow squares (red pentagons) corresponding to the measurements of $\langle{\sigma_z}\rangle$ and $\langle{N}\rangle$, respectively. 
Solid lines represent theoretical results including experimental imperfections (See in Supplement Material). The case with $u = 0 $ represents a static detector which observes no excitation and a vacuum photon field. The case with $u=1.5$ represents an oscillating detector. In the detector's frame, the photon field and the detector experience periodic excitation simultaneously, showing coordinated dynamics.  }
\label{fig:res}
\end{figure} 

As shown in Fig. \ref{fig:res}, when the detector stays stationary in the cavity ($u = 0$), there is no acceleration ($a(t)=0$), so the system remains at the ground state. For the oscillatory case with non-zero amplitude ($u\neq0$), both the detector excitation $\langle\sigma_z\rangle$ and photon number $\langle N \rangle$ experience a temporally correlated oscillation. This simulation shows an important aspect of the Unruh effect, that despite a vacuum for the photon field in the rest frame, in the view of a non-inertial detector, the photon is created jointly with the excitation of the detector. In contrast to the original model of Unruh effect, where a uniformly accelerated detector observes a thermal photon field, our observation of a Rabi-like oscillation indicates a strong coherent effect, given a resonant modulation of the coupling between the detector and the photon field.


\begin{figure}
\centering
\includegraphics[width=\columnwidth]{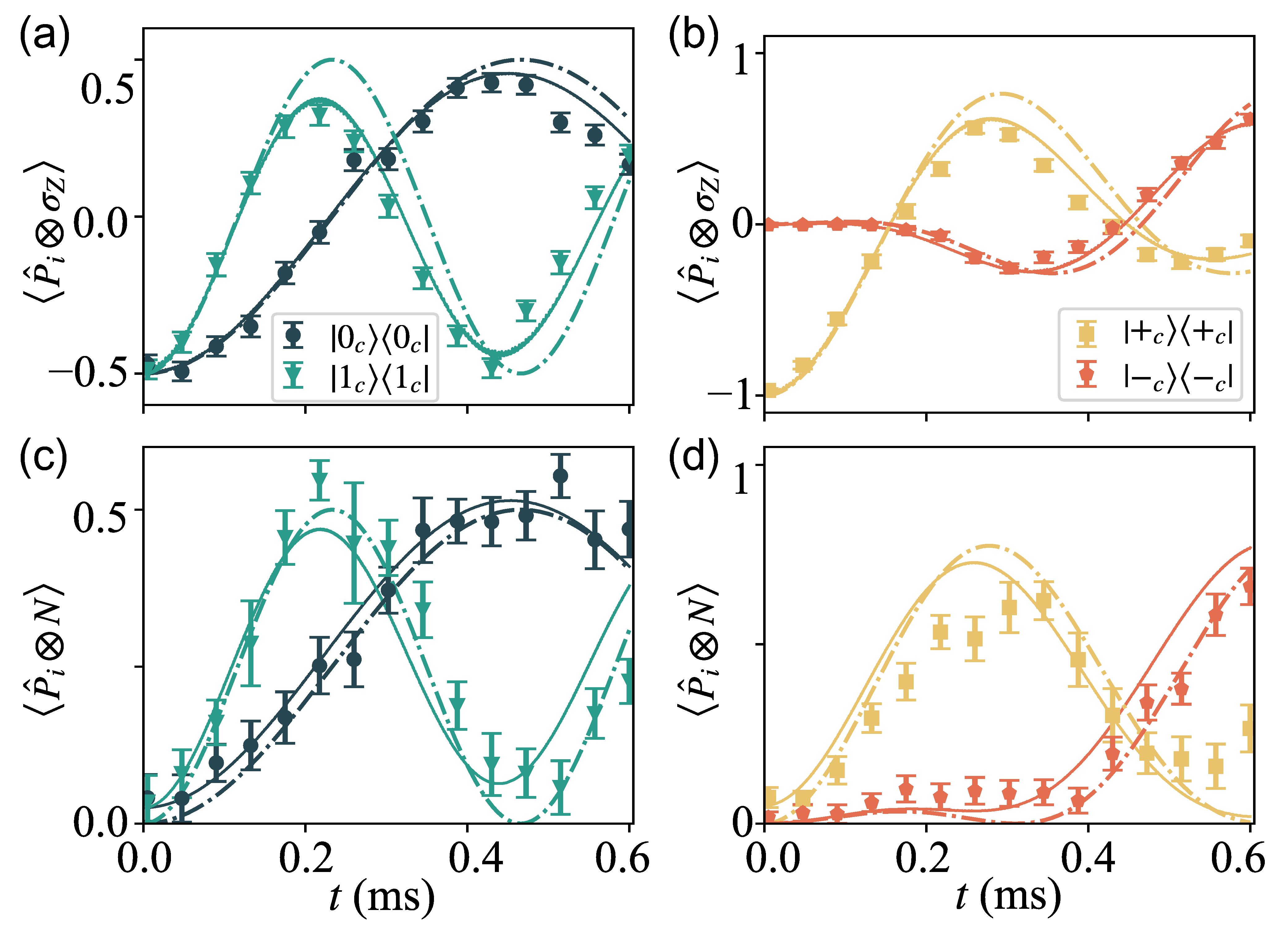}

\caption{Experimental simulation of superposed trajectories. (a) and (b) show the evolution of detector excitation $\langle\sigma_z\rangle$. (c) and (d) show the evolution of photon number $\langle N \rangle$. Solid (Dot-dashed) lines represent numerical simulation for a theoretical model with (without) experimental imperfections (See in Supplement Material). Data points with error bars represent the experimental results. Dark blue circles, green blue triangles, orange-yellow squares, and red-orange pentagons represent projectors $\hat{P}_i=\ketbra{i_c}$ with $i=\{0,1,+,-\}$, respectively. For the measurements regarding $\ketbra{0_c}$ and $\ketbra{1_c}$, both the detector and mean photon field excitation experience sinusoidal oscillation with different rates, due to distinguished trajectories of the detector. The difference between the experimental observation of $\langle\ketbra{+_c}\otimes\hat{O}\rangle$ and $\langle\ketbra{-_c}\otimes\hat{O}\rangle$ indicates a coherent superposition for the trajectories rather than a classical mixture, where such difference can be observed with both $\hat{O}=\{\sigma_z,N\}$. }
\label{fig:superposition}
\end{figure} 

\subsection{Experimental simulation of superposed trajectories}\label{sec5}

Here, we demonstrate a specific model with the superposition of relativistic trajectories. 

We still consider the resonant condition with $\omega = \omega_p + \omega_q$. The motion of the detector is set as a superposition of two trajectories $x_{i}(t)=L/2+A_{i}\sin(\omega t)$ of different amplitudes $A_{i}$, with $i=0,1$. Dimensionless amplitudes are denoted as $u_{i}= kA_{i}$.  The Hamiltonian for this system is shown in Eq. \eqref{Eq:sup_total}. 
Set $\omega_q=\omega_p=2\pi\times150$~kHz,  $g_0=2\pi\times 1.92$ kHz and different oscillation amplitudes $u_{0}=kA_{0}=0.5822$ and $u_{1}=kA_{1}=1.5$, leading to distinguishable trajectories of the detectors, corresponding to the respective evolution of the detector-photon system. 

The initial state is set as a superposed state $\ket{+_c}\ket{g,n=0}$. 
Thus, the state evolves as $\ket{\Psi(t)}$, shown in Eq.~\ref{eq:PsiT}. As depicted in Fig. \ref{fig:superposition} (a) and (c), we measure $\langle\hat{P}_i\otimes\hat{O}\rangle$ with $\hat{P}_i=\ketbra{i_c}$, $i=\{0,1\}$, and $\hat{O}=\{\sigma_z,N\}$ (same below). The joint oscillating excitation for both the detector and photon field is as expected for the single trajectory model mentioned above for each portion of the superposition. We also measure $\langle\hat{P}_i\otimes\hat{O}\rangle$ defined above for $i=\{+,-\}$ as depicted in Fig. \ref{fig:superposition} (b) and (d), where we experimentally observe highly distinguishable evolution curves between $\langle \ketbra{+_c}\otimes\hat{O}\rangle$ and $\langle \ketbra{-_c}\otimes\hat{O}\rangle$. Such difference can be explained by decomposing $\langle \ketbra{\pm_c}\otimes\hat{O}\rangle=(\langle\hat{O}\rangle_{\text{mix}}\pm\langle\hat{O}\rangle_{\text{coh}})/2$, where $\langle\hat{O}\rangle_{\text{mix}}=(\bra{\psi_0(t)}\hat{O}\ket{\psi_0(t)}+\bra{\psi_1(t)}\hat{O}\ket{\psi_1(t)})/2$ is a result of the incoherent mixture of both trajectories,  and the non-zero component $\langle\hat{O}\rangle_{\text{coh}}=(\bra{\psi_0(t)}\hat{O}\ket{\psi_1(t)}+\bra{\psi_1(t)}\hat{O}\ket{\psi_0(t)})/2$ is an indication of strong quantum coherence due to the superposed trajectories. Hence, our result shows the combined signatures of both relativistic effects by acceleration-induced radiation, and quantum-mechanical effects by coherent superposition.

\section{Conclusions and Discussions}\label{sec6}

In summary, we experimentally perform quantum simulation of 
acceleration-induced excitation for a detector in a cavity with coherently superposed trajectories. We simulate the detector and photon field with the spin and motion of a trapped ion, respectively. We apply coherently controlled laser couplings and observe vacuum excitation induced by non-inertial motion from an oscillating detector. We also explore the novel dynamical behavior of the detector and photon field, triggered by the coherent superposition of non-inertial trajectories. Our study can be generalized to explore a series of relativistic trajectories and related phenomena, including the Unruh effect with linear acceleration \cite{PhysRevD.14.870, RevModPhys.80.787} and Cherenkov radiation \cite{Cherenkov}. 

Meanwhile, our simulation could be readily scaled to more levels for internal qudit systems coupled with bosonic modes for various experimental systems including ions \cite{Nat.Phys..18.1053,QFT}, neutral atoms \cite{Phys.Rev.Appl..19.064060,Phys.Rev.Lett..129.160501}, photons \cite{NatCommun.13.1166} and superconducting circuits \cite{Phys.Rev.Lett..126.017702,Phys.Rev.B.97.094513}. 

Based on our demonstration, we consider a modification of a recent proposal \cite{PhysRevLett.125.241301}, where a particle of either an ion or an electron is trapped with microwave field \cite{electron}, and is coupled to a superconducting cavity. 
With an electric field configuration pushing the particle off the trap microwave field null, the particle would spatially oscillate at a frequency of the trap drive, providing the required $x(t)=A\cos(\omega t)$. Two internal states of the particle are set as a two-level detector with Zeeman or hyperfine splitting $\hbar \omega_q$ under an ambient magnetic field. 
In addition, the coupling superconducting cavity could provide a high-quality factor $Q=10^7$ and small mode volume $V=10^{-14}\text{ m}^3$ at mode frequency $\omega_p$. Thus, the Unruh effect could be enhanced under resonant condition $\omega_p=\omega-\omega_q$, as verified in this work. Assuming a motion frequency of $\omega =2\pi\times5$~GHz, an energy spacing of $\omega_q=2\pi\times50$~MHz \cite{electron, Nature.645.362} and a micromotion amplitude of $5~$\textmu m, we expect a joint excitation of the spin and motion to occur with $0.2$~mHz rate within the spin life time, while no excitation is expected for inertial motion. Such an experiment is currently under investigation.

~~~~~
\begin{acknowledgments}
We thank Mengxiang Zhang and other teammates for the serious discussion of experimental details. We acknowledge support from the National Natural Science Foundation of China (grant number 92165206) and Innovation Program for Quantum Science and Technology (Grant No. 2021ZD0301603). Zehua Tian was supported by the scientific research start-up funds of Hangzhou Normal University: 4245C50224204016. 

\textbf{Conflict of Interest  } The authors declare that they have no conflict of interest.
\end{acknowledgments}




%



\pagebreak
\clearpage
\widetext

\begin{center}
\textbf{\large Supplementary Material}
\end{center}

\setcounter{equation}{0}
\setcounter{section}{0}
\setcounter{figure}{0}
\setcounter{page}{1}
\makeatletter
\renewcommand{\thefigure}{S\arabic{figure}}  
\renewcommand{\theequation}{S\arabic{equation}} 
\section{Implementation of the model}\label{Sec:SimImp}

\textbf{State Preparation:} The initial state is prepared to $\ket{S_{1/2},+1/2}$ through optical pumping, then apply $R_{1/2,1/2}(\pi/2,\pi/2)$ and $R_{-1/2,1/2}(\pi,3\pi/2)$ pulses sequentially. Here $R_{i,j}(\theta,\phi)$ denotes a resonant pulse driving transition $\ket{S_{1/2},i}\leftrightarrow\ket{D_{5/2},j}$ with rotation angle of $\theta$ and azimuthal angle of $\phi$ on the Bloch sphere, with $i=\{-1/2,1/2\}$ and $j=\{-5/2,-3/2,-1/2,1/2,3/2,5/2\}$. Then the initial state is prepared to $\ket{n=0}(\ket{S_{1/2},+1/2}+\ket{S_{1/2},-1/2})/\sqrt2$, which is $\ket{+_c}\ket{g,0}$ in the model. All these resonant pulses are discernible in frequency domain at $\sim10$~MHz level due to the fine splitting under magnetic field. So multiple couplings could be individually driven, modulated and applied simultaneously. The frequencies of the vibration modes of the ion $\omega_r/2\pi$ are $0.88$~MHz for single trajectory experiment and $1.12$~MHz for superposed trajectories experiment. By picking laser detuning at $\pm\omega_r$, one could couple vibration mode and internal levels of the ion (sideband transitions). We utilize sidebands to perform ground state cooling of the vibrational modes.

\textbf{Detector-Photon Interaction:} 
As defined in the main text, the target Hamiltonian is
\begin{equation}\label{Eq:model}
\begin{aligned}
    H =\hbar\omega_p a^\dagger a + \cfrac{\hbar\omega_q}{2}\sigma_z
     + \hbar g_0 \sin(k x(t))\sigma_x(a+a^\dagger).
\end{aligned}
\end{equation}

Start with trapped ion Hamiltonian under laser driving. 
\begin{equation}\label{Eq:ionSB}
\begin{aligned}
    H_{\text{ion}}+H_{\text{Couple}} =\hbar\omega_r a^\dagger a + \cfrac{\hbar\omega_0}{2}\sigma_z
     + \hbar g_0 \sigma_x \{e^{i\left[\eta(a+a^\dagger)-\omega_{L}t+\phi\right]}+\text{H.c.}\},
\end{aligned}
\end{equation}
where $\omega_0$ is the energy difference of internal levels, which is near the electric quadruple driving 729 nm laser photon energy. Dimensionless parameter $\eta$ is the Lamb-Dicke parameter \cite{J.Res.Natl.Inst.Stand.Technol..103.259}. Move into the interaction picture with $H_0=\hbar(\omega_r-\omega_p) a^\dagger a + \cfrac{\hbar(\omega_0-\omega_q)}{2}\sigma_z$, then phonon and internal levels shift to $H_{\text{ion}}^{(I)} =\hbar\omega_p a^\dagger a + \cfrac{\hbar\omega_q}{2}\sigma_z$, which matches the photon and detector in Eq. \eqref{Eq:model}. 
Furthermore, by picking laser detuning at $\omega_L-\omega_0=-\omega_q+(\omega_r-\omega_p)$ with phase $\phi_B$, we could derive the off-resonant blue sideband $H_{\text{Couple}}^{(I)} \approx \hbar g_0 [i \eta\sigma_+ e^{i(\omega_0-\omega_q)t}a^{\dagger}e^{i(\omega_r-\omega_p)t}e^{-i\omega_{L}t+i\phi_B}+\text{H.c.}]$. Similarly, off-resonant red sideband is achieved with $\omega_L-\omega_0=-\omega_q-(\omega_r-\omega_p)$ and phase $\phi_R$. 
Combing red and blue off-resonant sidebands \cite{Nat.Phys..16.1206}, one could obtain
\begin{equation}\label{Eq:exp model}
\begin{aligned}
    H =&\hbar\omega_p a^\dagger a + \cfrac{\hbar\omega_q}{2}\sigma_z + \hbar g_0 
    \times
    [-\sin(\phi_{+})\sigma_x
    -\cos(\phi_{+})\sigma_y]
    \times
    [\cos(\phi_{-})(a+a^\dagger)
    +i\sin(\phi_{-})(a-a^\dagger)],
\end{aligned}
\end{equation}
where $\phi_{\pm}=(\phi_{\text{R}}\pm\phi_{\text{B}})/2$. We set $\phi_{\text{R}}=\phi_{\text{B}}=3\pi/2$, then Eq. \eqref{Eq:exp model} matches Eq. \eqref{Eq:model} after further temporal modulation of $g_0$.
In this paper, we set $\omega_p=\omega_q=2\pi\times150\text{ kHz}$. The strength of both pulses is proportional to $\sin(kx(t))$, which is modulated with arbitrary waveform generators (AWG) to simulate the modulation due to relativistic trajectories. This programmable modulation enables quantum simulation of other trajectories.

For the implementation of superposed trajectories in the main text, the effective Hamiltonian could be derived similarly with the implementation of the single trajectory model. The resonant frequencies transitions of $\ket{S_{1/2},+1/2}\leftrightarrow\ket{D_{5/2},+1/2}$ and $\ket{S_{1/2},-1/2}\leftrightarrow\ket{D_{5/2},-1/2}$ are separated by $8$ MHz. Thus, sideband pulses could individually drive corresponding interaction terms. Then two pairs of detuned red and blue sideband pulses are applied at the same time, with modulation strengths of $\sin(kx_{0}(t))$ and $\sin(kx_{1}(t))$, respectively. Following Eq. \eqref{Eq:exp model}, each pair of sideband pulses introduces interaction form $\ketbra{i_c}\otimes\sigma_x(a+a^\dagger)$. The corresponding Hamiltonian becomes the form in the main text:
\begin{equation}\label{Eq:Sup}
\begin{aligned}
    H =\hbar\omega_p a^\dagger a + \cfrac{\hbar\omega_q}{2}\sigma_z
    + \sum_{i=0,1} \hbar g_0 \sin(k x_i(t))\ket{i_c}\bra{i_c}\otimes\sigma_x(a+a^\dagger).
\end{aligned}
\end{equation}

\textbf{Experimental Imperfections: }{Experimental imperfections are quantum decoherence and imperfect initial state preparation. The major decoherence terms are Lindblad form $\cfrac{{\rm d}\rho}{{\rm d}t}=-\cfrac{i}{\hbar}[H,\rho]+\sum_i\gamma_i\left(L_i\rho L_i^\dagger-\cfrac{1}{2}\{L_i^\dagger L_i,\rho\}\right)$ with damping rates $\gamma_i$ and jump operators $L_i$. Jump operators are $\sigma_z$ for dephasing and $a$ and $a^{\dagger}$ for phonon heating rate. The damping rates reflect in experimental metrics. We measure them through additional calibration experiments. Dephasing is characterized with the coherence time $T_2$ in a Ramsey experiment; the heating rate $\dot{N}$ is characterized by measuring average phonon number $\bar{N}$ after variable waiting time and fitting $\bar{N}\sim T_{\text{wait}}$. 
During the single trajectory experiment $T_2=1$ ms  and heating rate $\dot{N}=20$ quanta/s are measured. }
{During the superposed trajectories experiment $T_2=1$ ms  and heating rate $50$ quanta/s are measured. Since four fine states are involved here, $\ket{S_{1/2},1/2}\to\ket{D_{5/2},1/2}$ transition is set to calibrate $T_2$. The dephasing strength of each state is inferred with their sensitivity of the magnetic field. For imperfect initial state preparation mainly comes from the vibrational modes. The initial ground state of vibrational mode is prepared with a cooling sequence, which leaving a little amount of excitation beyond $|n=0\rangle$. Experimental measurement results shows a residual thermal state of $\bar{N}=0.05$.}
These experimental imperfections cause the experimental data to deviate slightly from the theoretical predictions, but do not affect the main conclusions of the experiment. 

\section{Model analysis}\label{Sec:ModelAnalysis}

For oscillatory trajectory form, Floquet analysis could be performed to derive the evolution analytically. The time-dependent coupling strength in Eq. (\ref{Eq:model}) could be expanded with Jacobi-Anger expansion,
\begin{equation}\label{Eq:JA}
\begin{aligned}
    &\sin(k \bar{x} + k A\sin(\omega t)):=\sin(\bar{u} + u\sin(\omega t))\\
    =&\sin (\bar{u})[J_0(u)+2\sum_{n=1}^\infty J_{2n}(u)\cos(2n\omega t)]
    +2\cos (\bar{u}) \sum_{n=1}^\infty J_{2n-1}(u)\sin((2n-1)\omega t).
\end{aligned}
\end{equation}
Here $J_n(u)$ is the Bessel functions. Simple combination of Jaynes-Cummings(JC) and anti-Jaynes-Cummings(anti-JC) terms could be derived by further transformation into the interaction picture with $H_0=\hbar\omega_p a^\dagger a + \cfrac{\hbar\omega_q}{2}\sigma_z$  and keeping the resonant terms only \cite{Phys.Rev.A.92.033817,Phys.Rev.A.96.032121}. Take our set in the main text as an example: $\omega=\omega_p+\omega_q$, $\omega_p=\omega_q$ and $\bar{u}=0$. 
\begin{equation}\label{Eq:res}
\begin{aligned}
    H &=\hbar\omega_p a^\dagger a 
    + \cfrac{\hbar\omega_q}{2}\sigma_z
    + \hbar g_0 \sin(k x(t))\sigma_x(a+a^\dagger)\\
    \stackrel{H_0}{\longrightarrow}\quad 
    H_{\text{Int}}/\hbar g_0&=
    \sin(\bar{u} + u\sin(\omega t))
    \left[
    (\sigma^+ae^{i(\omega_q-\omega_p)t}+\sigma^-a^{\dagger}e^{-i(\omega_q-\omega_p)t})
    +(\sigma^+a^{\dagger}e^{i(\omega_q+\omega_p)t}+\sigma^-ae^{-i(\omega_q+\omega_p)t})
    \right]\\
    &=\sin (\bar{u})J_0(u)(\sigma^+a+\sigma^-a^{\dagger})
    +2\cos (\bar{u}) J_{1}(u)\sin(\omega t)(\sigma^+a^{\dagger}e^{i(\omega_q+\omega_p)t}+\sigma^-ae^{-i(\omega_q+\omega_p)t})+\cdots\\
    &=-i J_{1}(u)(e^{i(\omega_q+\omega_p)t}-e^{-i(\omega_q+\omega_p)t})(\sigma^+a^{\dagger}e^{i(\omega_q+\omega_p)t}+\sigma^-ae^{-i(\omega_q+\omega_p)t})+\cdots\\
    &=i J_{1}(u)(\sigma^+a^{\dagger}-\sigma^-a)+\cdots, 
\end{aligned}
\end{equation}
where off-resonant terms are all omitted. The model Hamiltonian evolves under the anti-Jaynes-Cummings form $\sigma_+a^\dagger+\sigma_- a$ with strength $g_0J_1(u)$. With initial state $|g,n=0\rangle$, the evolution acts as Rabi-oscillation between $|g,n=0\rangle$ and $|e,n=1\rangle$, which exactly matches the experimental results in the main text.
As for the trajectory superposition model, the evolution of $\ket{\psi_0(t)}$ and $\ket{\psi_1(t)}$ could be evaluated in the same way. The final state could be combined with control qubit as $\frac{1}{\sqrt{2}}\left(\ket{0_c}\ket{\psi_0(t)}
+\ket{1_c}\ket{\psi_1(t)}\right)$.

Effective interaction Hamiltonian of oscillatory trajectories under different conditions could be estimated as well. For example, keeping $\omega=\omega_q+\omega_p$ and $\omega_q=\omega_p$ but altering $\bar{u}$ and $u$, we have
\begin{equation}\label{Eq:res_all}
\begin{aligned}
    H_{\text{Int}}/\hbar g_0&=\sin (\bar{u})J_0(u)(\sigma^+a+\sigma^-a^{\dagger})
    +i \cos(\bar{u})J_{1}(u)(\sigma^+a^{\dagger}-\sigma^-a)+\cdots.
\end{aligned}
\end{equation}
Generally, the effective Hamiltonian is the combination of JC and anti-JC terms. Only when anti-JC term is non-zero, the system would endure excitation from ground state, which means $\cos(\bar{u})\not=0$ and $J_1(u)\not=0$. So the Unruh-type excitation occurs only when the acceleration is non-zero ($kA=u\not=0\to A\not=0$).

Another interesting condition discussed in \cite{Phys.Rev.B.92.064501} is $\omega=\omega_q=\omega_p$, where
\begin{equation}\label{Eq:cre}
\begin{aligned}
    H_{\text{Int}}/\hbar g_0&=\sin (\bar{u})J_0(u)(\sigma^+a+\sigma^-a^{\dagger})
    + \sin(\bar{u})J_{2}(u)(\sigma^+a^{\dagger}+\sigma^-a)+\cdots.
\end{aligned}
\end{equation}
Further picking $J_0(u)=J_2(u)$ leads to spin-dependent displacement on the photon as $\ket{g,0}\to\ket{+}\ket{\alpha}+\ket{-}\ket{-\alpha}$, which might contribute considerable excitation in direct realization scheme.

This model could also investigate uniform acceleration trajectory within the cavity. However, as discussed in \cite{Phys.Rev.A.102.042223}, significant excitation occurs only when the velocity exceeds the Cherenkov threshold, which implies surpassing the speed of light in the medium and achieving an extremely long travel distance. So uniform acceleration is physically difficult to realize.

\section{Measurement process}\label{Sec:Meas}
\begin{figure}
\centering
\includegraphics[width=0.9\textwidth]{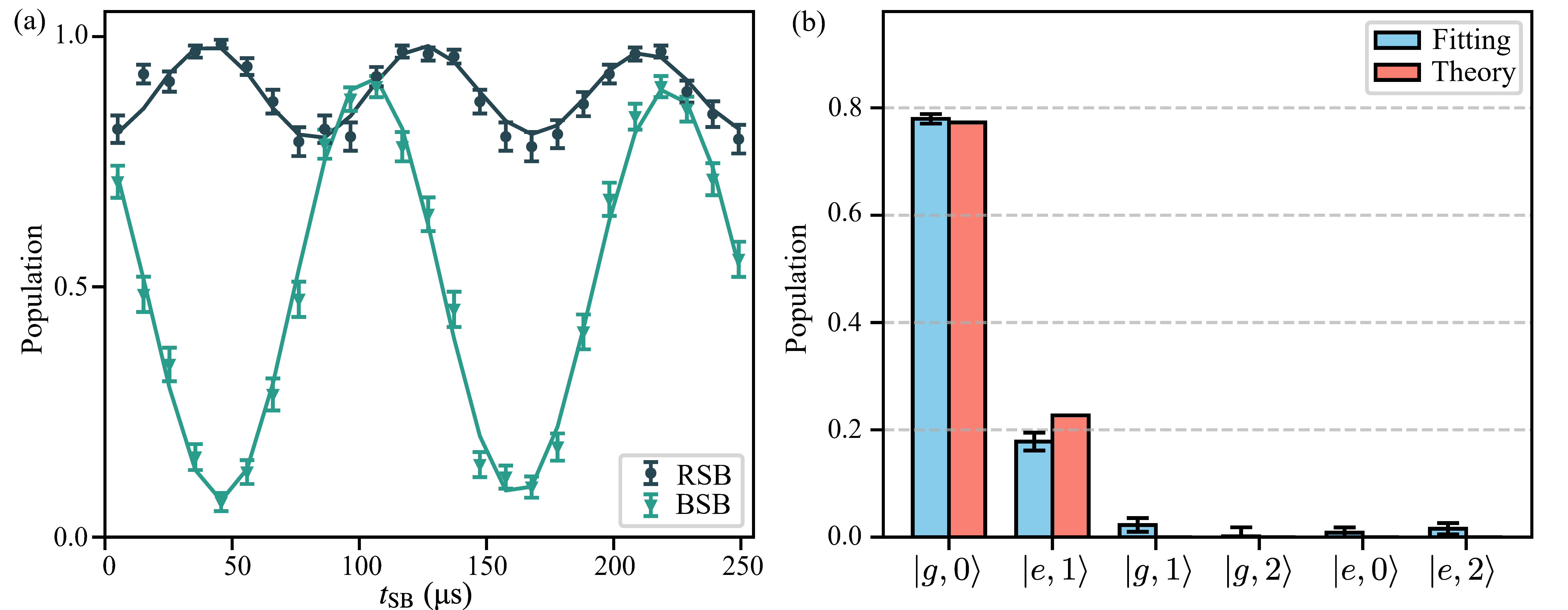}
\caption{(a) Experiment data and fitting curves of scanning red and blue sideband pulse length $t_{\text{SB}}$ after evolution. (b) Fitted phonon distribution. Data points with error bars are experimental results. Solid curves represent fit results. The dark and cyan color stands for red sideband and blue sideband. Both main oscillation parts of the red/blue sideband curve could be found. Combining estimation of Lamb-Dicke parameter $\eta=0.065$, red/blue sideband oscillation part could be distinguished as $\ket{e,n=1}\to\ket{g,n=2}$ and $\ket{g,n=0}\to\ket{e,n=1}$ transition. The red sideband oscillates faster than the blue sideband. The fitting result is $\langle N \rangle=0.235\pm0.045$. The phonon mostly distributed in $\ket{g,0}$ and $\ket{e,1}$, which match theoretical expectation.}
\label{fig:fit}
\end{figure}

For simulation with the single trajectory, to measure $\langle\sigma_z\rangle$, we detect the population of $\ket{g}$ (i.e. $\ket{S_{1/2},+1/2}$) after evolution, denoted as $p_{g}$. Then we have $\langle\sigma_z\rangle=1-2p_{g}$. To measure $\langle N\rangle$, we individually apply the red and blue sideband pulses after evolution. The sideband pulse length $t_{\text{SB}}$ is scanned up to $250$ \textmu s. The phonon population could be derived from sideband scan due to Rabi frequency's dependence of phonon number \cite{J.Res.Natl.Inst.Stand.Technol..103.259}. 
For example, blue sideband couples $\ket{g,n}$ and $\ket{e,n+1}$. The probability amplitude of $\ket{g,n}$ after evolution time $t$ is $c_{g,n}(t)=\cos(\Omega_{n,n+1}t/2)c_{g,n}(0) +\sin(\Omega_{n,n+1}t/2)c_{e,n+1}(0)$, where $\Omega_{n,n+1}=\eta\Omega_0 e^{-\eta^2/2}L_{n}^{1}(\eta^2)/\sqrt{n+1}$ is the Rabi frequency, and $L_{n}^{1}$ is generalized Laguerre polynomials. We denote the population on each state as $P_{g/e,n}(t)=|c_{g/e,n}(t)|^2$, where we are interested in extracting $P_{g,n}(0)$ and $P_{e,n}(0)$ before the application of sideband measurements. The total population on $\ket{g}$ is $P_g(t)=\sum_{n}[P_{n+} +P_{n-}\cos(\Omega_{n,n+1}t)+S_{n}\sin(\Omega_{n,n+1} t)]/2$, where $P_{n+}=P_{g,n}(0)+P_{e,n+1}(0)$, $P_{n-}=P_{g,n}(0)-P_{e,n+1}(0)$ and $S_n=c_{g,n}^*(0)c_{e,n+1}(0)+c_{g,n}(0)c_{e,n+1}^*(0)$. So the blue sideband scan gives evaluation for $\sum_{n}[P_{g,n}(0)+P_{e,n+1}(0)]/2$ and each $P_{n-}=[P_{g,n}(0)-P_{e,n+1}(0)]/2$. Similarly, the red sideband scan gives evaluation for $P_{g,0}+\sum_{n}[P_{g,n}(0)+P_{e,n-1}(0)]/2$ and each $P^\prime_{n-}=[P_{g,n}(0)-P_{e,n-1}(0)]/2$. As a result, a combined fitting of blue and red sideband scans could derive all $P_{g,n}(0)$ and $P_{e,n}(0)$. So the average photon number could be deduced with $\langle\ketbra{i}\otimes N\rangle=\sum_n nP_{i,n}$, where $i=\{g,e\}$. A typical fitting result is shown in Fig. \ref{fig:fit}, giving $\langle N \rangle=0.235\pm0.045$. 

For the simulation of superposed trajectories, the measurement process is extended to four levels. After evolution, there exists non-zero population on four states $\ket{S_{1/2},\pm1/2},\ket{D_{5/2},\pm1/2}$ ($\ket{0_c/1_c,g},\ket{0_c/1_c,e}$ in the model). Here projectors $\hat P_{i}=\ketbra{i_c}$ with $i= \{0,1,+,-\}$ are defined. The population values could derive the projected detector excitation with $\langle\hat{P}_{i}\otimes\sigma_z\rangle=p_{i_c,e}-p_{i_c,g}$, where $p_{i_c,e(g)}$ denotes population value of $\ket{i_c,e(g)}$, with $i=\{0,1,+,-\}$. During the detection process, both $\ket{S_{1/2},-1/2}$ and $\ket{S_{1/2},+1/2}$ population contributes to the bright counts. To obtain the population of one of the states, the population of the other state is moved to $\ket{D_{5/2}}$ with a $\pi$ pulse. For example, after a $R_{-1/2,-5/2}(\pi,\pi/2)$ pulse, the detection process gives population on $\ket{S_{1/2},+1/2}$. We call this process of removing unwanted populations before detection as shelving. So $p_{0_c(1_c),g}$ could be measured individually with the respective shelving process. 
To obtain population on $\ket{D_{5/2,\pm1/2}}$, $R_{\pm1/2,\pm1/2}(\pi,\pi/2)$ pulse is applied to swap population of $\ket{S_{1/2,\pm1/2}}$ and $\ket{D_{5/2,\pm1/2}}$. Then the original $p_{0_c(1_c),e}$ could be measured with detection after shelving. Measurement of $\langle\hat{P}_{0(1)}\otimes N\rangle$ is performed after shelving away the unwanted population on $\ket{{1_c(0_c),g}}$, using the same method in single trajectory simulation. To obtain expectation value $\langle\hat{P}_{\pm}\otimes\sigma_z\rangle, \langle\hat{P}_{\pm}\otimes N\rangle$, a transform pulse series is performed to achieve $\ket{+_c}\to\ket{0_c}$ and $\ket{-_c}\to-\ket{1_c}$. Then the following process is the same as our discussion above. Specifically, it is implemented with a pulse sequence  of $R_{-1/2,1/2}(\pi,\pi/2)-R_{1/2,1/2}(\pi/2,3\pi/2)-R_{-1/2,-1/2}(\pi/2,\pi/2)-R_{-1/2,1/2}(\pi,3\pi/2)$.

\end{document}